\documentstyle[floats,twocolumn,aps,prl,epsf]{revtex}
\begin{document}
\ifpreprintsty\else
\twocolumn[\hsize\textwidth%
\columnwidth\hsize\csname@twocolumnfalse\endcsname
\fi
\draft
\tighten
\title{Electron Dynamics in a DNA Molecule}
\author{V.M. Apalkov$^\dagger$ and
Tapash Chakraborty$^{\ast\ddag}$}
\address{$^\dagger$Department of Physics and Astronomy, 
Georgia State University, Atlanta,
Georgia 30303, USA
}
\address{$^\ast$Department of Physics and Astronomy,
The University of Manitoba, Winnipeg, Canada R3T 2N2}
\date{\today}
\maketitle
\begin{abstract}
We report on our theoretical investigations of the electronic
states in a DNA molecule. We have used a two-leg charge ladder model
where electron-electron interactions and the electron spin
have been taken into account. The energy spectra for G-C and
A-T base pairs obtained by numerically diagonalizing the Hamiltonian
reveal a gap structure and the interaction is found to
enhance the energy gaps. We also present the charge distribution
in the ground state and low-lying excited states for the A-T and G-C
base pairs. 
\end{abstract}
\ifpreprintsty\clearpage\else\vskip1pc]\fi
%\pacs{73.21.-b,87.14.Gg, 85.65.+h}
\vskip2pc

\narrowtext

The unique properties of DNA, self-assembly and molecular 
recognition, has rendered the `molecule of life' \cite{books} 
a promising candidate in the rapidly emerging field of molecular 
nano-electronics \cite{reviews}. A recent report of a field-effect 
transistor based on DNA molecules \cite{yoo}, that was preceded 
by a series of seminal experiments on the electron conduction
in DNA \cite{metallic}, has sparked a lot of interest on the 
electronic properties of the DNA. A thorough understanding of
the electronic properties of DNA is crucial in the development of
the future DNA-based nanoscale devices. In addition, charge transfer 
through DNA also plays an important role in radiation damage and 
repair \cite{giese} and therefore important for biological processes. 
Following the techniques pioneered earlier for a direct measurement 
of electrical transport through a DNA molecule \cite{porath},
Yoo et al. \cite{yoo} measured the conductance of poly(dG)-poly(dC) 
DNA (1.7 - 2.9 $\mu$m long) and poly(dA)-poly(dT) DNA (0.5 - 
1.5 $\mu$m). The bundle of DNA molecules were trapped between 
electrodes that were 20 nm apart. The I-V results showed a strong 
temperature dependence of the current and was interpreted in terms 
of a small polaron hopping model where the current
was expressed as $I\propto \sinh \beta V\,{\rm e}^{-E_a/kT}$. Here,
$E_a$ is the activation energy, $T$ is the background temperature,
$\beta=ea/2kT d$, $e$ is the electron charge, $a$ is the hopping 
distance, and $d$ is the separation between the electrodes. The 
results for poly(dA)-poly(dT) DNA can very well be described in 
this way if $\beta$ is taken to be independent of temperature. 
In the case of poly(dG)-poly(dC) DNA molecules, a similar temperature 
dependence was observed but with a much lower resistance of 1.3 
M$\Omega$ at room temperature as compared to 100 M$\Omega$ for 
the poly(dA)-poly(dT) DNA. The poly(dG)-poly(dC) DNA
also shows the correct temperature dependence of $\beta$ and the
temperature dependence of the current was observed down to 4.2 K. In
contrast, for the poly(dA)-poly(dT) DNA, temperature dependence of the
current was observed down only to 50 K.

Yoo et al. also performed I-V measurements with an additional electric
field from the back of the Si substrate and measured the FET property
of the back-gate type. They noted that, in the FET based on 
poly(dA)-poly(dT), application of a positive gate voltage results in 
a larger conductance, i.e., an indication of $n$-type behavior. In 
the case of the DNA-FET based on poly(dG)-poly(dC), a $p$-type 
conducting behavior was observed.

The DNA conductivity measurements discussed above have also inspired 
a few theoretical studies 
\cite{yi,asai,brunaud,cuniberti,iguchi,kats,revmod,intrinsic}. 
These works primarily focused on the evaluation of transport through 
one-dimensional systems using a model Hamiltonian, or electronic 
structure calculations from first principles. It should be pointed
out that the effect of electron-electron interactions is important for 
understanding the physical properties of the DNA molecules, such as 
excitation spectra, charge distribution and charge transport in DNA molecules. 
The reason for that is the following: There are a few energy scales which 
determine the DNA properties. The first one is the tunneling gap, or hopping
integrals between nearest DNA base pairs. The typical value of these 
hopping integrals is $0.5$ eV. The second energy scale is determined by 
the single-particle energy spectrum of a single DNA base pair. This is the
energy between the highest occupied molecular orbital (HOMO) and the lowest 
unoccupied molecular orbital (LUMO), which is of the order of  1 eV \cite{yi}. 
The other energy scale comes from interactions between electrons 
at different and at the same base pairs. This energy scale is $1$ eV
for Hartree interactions and since it is close to HOMO to LUMO 
excitation energy the interaction can have strong effect on 
many-particle excitation spectra and charge distribution. At the same
time the exchange interaction, which can be of the order of hopping 
integrals between nearest base pairs, should have strong effect on
electron transport.    

In this paper, we report on the electronic properties of the DNA, in 
particular, the influence of electron-electron interaction on the energy 
spectrum and the excitation gap. We model the double-stranded DNA as a 
two-leg charge ladder \cite{yi}. As a first approximation, we consider 
only the electronic degree of freedom and disregard the vibrational 
modes and their effects on the electronic motion \cite{vibration}. The 
Hamiltonian of the electronic system consists of two parts: the tight-binding 
Hamiltonian, ${\cal H}_t$, and interaction Hamiltonian, ${\cal H}_i$. The 
tight-binding Hamiltonian is a two-chain Anderson Hamiltonian describing 
the hopping between the nearest neighbors (nearest bases) within each 
chain and the hopping between the two chains (within each base)

\begin{eqnarray}
{\cal H}_{t} & = & \sum_{i\sigma} \varepsilon_{h} a^{\dagger }_{i,\sigma} 
a_{i,\sigma}+\sum_{i\sigma}\varepsilon_{l} b^{\dagger }_{i,\sigma} 
b_{i,\sigma} + \nonumber \\
& & \sum_{i\sigma} t_{h} \left[ a^{\dagger }_{i,\sigma} a_{i+1,\sigma}+ 
h.c. \right]+ \sum_{i\sigma} t_{l} \left[ b^{\dagger }_{i,\sigma}
 b_{i+1,\sigma}+ h.c. \right]
\nonumber \\
& & + \sum _{i\sigma} t_{hl}\left[ a^{\dagger }_{i,\sigma} b_{i,\sigma}+ 
h.c. \right]
\label{Ht}
\end{eqnarray}
where $\varepsilon_h$, $\varepsilon_l$ are the (site) energies of 
the HOMO and LUMO for a single isolated base pair respectively, 
$a_{i,\sigma}$, $b_{i,\sigma}$ are annihilation operators of electron 
with spin $\sigma$ in HOMO and LUMO states of $i$-th base pair, $t_h$
is the hopping integral between HOMO of the nearest base pairs, $t_l$ is 
the hopping integral between LUMO of the nearest base pairs, and $t_{hl}$ is
the hopping integral between strands (HOMO and LUMO) of the same base pair. 
The index $i$ labels the different base pairs, while $\sigma=\uparrow\downarrow$ 
is the spin index. The summation over index $i$ goes from 1 to $N$ 
where $N$ is the number of base pairs.

The interaction part of the Hamiltonian has the following form
\begin{eqnarray}
& & {\cal H}_i=\sum_{i\sigma}V_{h0}\left(a^\dagger_{i,\sigma} a_{i,\sigma}
               \right) \left(a^{\dagger}_{i,-\sigma} a_{i,-\sigma}\right) 
              \nonumber \\
 & & + \sum_{i\sigma}V_{l0}\left(b^{\dagger}_{i,\sigma} b_{i,\sigma}\right)
              \left(b^{\dagger}_{i,-\sigma} b_{i,-\sigma} \right)
        \nonumber \\
 & & + \sum_{i\sigma\sigma_1}V_{hl0}\left(a^{\dagger}_{i,\sigma} a_{i,\sigma}\right)
              \left(b^{\dagger}_{i,\sigma_1} b_{i,\sigma_1} \right) 
             \nonumber \\
  & & - \sum_{i}V^{(ex)}_{hl0} \left(a^{\dagger}_{i,\sigma} a_{i,\sigma}\right)
             \left(b^{\dagger}_{i,\sigma} b_{i,\sigma} \right)
     \nonumber \\
      & & +   \sum_{i\sigma\sigma_1}
    V_{h1}\left(a^{\dagger}_{i,\sigma} a_{i,\sigma}\right)
          \left(a^{\dagger}_{i+1,\sigma_1} a_{i+1,\sigma_1} \right) 
    \nonumber \\
  & & -\sum_{i\sigma}V^{(ex)}_{h1}\left(a^{\dagger}_{i,\sigma} a_{i,\sigma}\right)
       \left(a^{\dagger}_{i+1,\sigma} a_{i+1,\sigma} \right)
      \nonumber \\
      & & +   \sum_{i\sigma\sigma_1}
         V_{l1} \left(b^{\dagger}_{i,\sigma} b_{i,\sigma}\right)
               \left(b^{\dagger}_{i+1,\sigma_1} b_{i+1,\sigma_1} \right)
    \nonumber \\
  & & - \sum_{i\sigma}V^{(ex)}_{l1} \left(b^{\dagger}_{i,\sigma} b_{i,\sigma}\right)
               \left(b^{\dagger}_{i+1,\sigma} b_{i+1,\sigma} \right)
       \nonumber \\
      & & +   \sum_{i\sigma\sigma_1}
        V_{hl0} \left(a^{\dagger}_{i,\sigma} a_{i,\sigma}\right)
          \left(b^{\dagger}_{i+1,\sigma_1} b_{i+1,\sigma_1} \right) 
    \nonumber \\
  & & - \sum _{i\sigma}V^{(ex)}_{hl0} \left( a^{\dagger }_{i,\sigma} a_{i,\sigma}
        \right)\left( b^{\dagger }_{i+1,\sigma} b_{i+1,\sigma} \right),
\label{Hi}
\end{eqnarray}
which is described by ten parameters.

In the ground state the number of electrons is equal to $2N$, so that all
HOMO states (with both spin directions) are occupied. To find the
excitation gap and the energy spectrum of the electron system with the
Hamiltonian Eqs.~(\ref{Ht},\ref{Hi}), we consider the DNA structure to have a 
finite number $N$ of base pairs and, by exactly diagonalizing the Hamiltonian
matrix, we obtain the ground state and the lowest excitation states of the system.
Some of the parameters of the DNA structure relevant for our studies are 
listed in Table I. To eliminate the effects of boundaries we have also imposed
the periodic boundary conditions, so that in Eqs.~(1)-(2) we have
$a_{N+1}=a_1$ and $b_{N+1}=b_1$.

\begin{table}
\caption{\label{tab:table} Paramaters of the DNA structure
used in our present work. Energies are in eV. The subscripts $h$ and 
$l$ correspond to the HOMO and LUMO.
}
\begin{tabular}{l|l|r}
Hopping integrals \\
\cite{endres,note1} & G-C & $t_h=-0.1419$, $t_l=0.0525$
\\ \cline{2-3}
& A-T & $t_h=-0.0695$, $t_l=0.1054$ \\ \hline
The site energies 
\cite{brunaud} &
G-C & $\varepsilon_h=-14.714$, $\varepsilon_l=-13.303$ \\ \cline{2-3}
& A-T & $\varepsilon_h=-14.635$, $\varepsilon_l=-13.734$ \\  \hline
On-site \\
interaction\cite{note2} & G-C & $V_{h0}=5.879$, $V_{l0}=5.227$ \\ \cline{2-3}
& A-T& $V_{h0}=5.681$, $V_{l0}=5.23$ \\ \hline
Interstrand \\
interaction \cite{brunaud} & G-C & $V_{h1}=1.844$, $V_{l1}=2.455$ \\ \cline{2-3}
& A-T& $V_{h1}=1.625$, $V_{l1}=2.378$ \\ \hline
Intrastrand \\
interaction \cite{brunaud}& G-C & $V_{hl0}=2.7$ \\ \cline{2-3}
& A-T& $V_{hl0}=2.6$ \\ 
\end{tabular}
\end{table}

We have considered a segment of DNA containing ten base pairs of 
poly(dG)-poly(dC) and poly(dA)-poly(dT) DNA molecules.
It should be pointed out that in the experiments of Porath et
al. \cite{porath}, the DNA sample has 30 base pairs. Similarly, we 
can estimate that since the distance between electrodes in the 
experiments by Yoo {\em et al.} \cite{yoo} was about 20 nm, taking 
into account the fact that the distance between the base pairs is 
$0.34$ nm, there were $\sim 50$ base pairs between electrodes of 
the set up used by Yoo et al. However, in our present model, 
handling more than 10 base pairs would be a formidable endeavor
and has not been attempted.

\begin{figure}
\begin{center}
\begin{picture}(120,130)
\put(0,0){\includegraphics{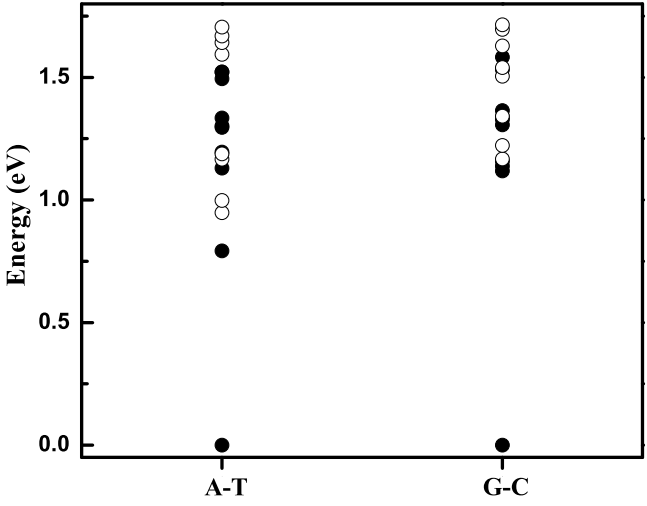}}
\end{picture}
\vspace*{2.0cm}
\caption{The energy spectra of the A-T and the G-C base pairs
in our model. The closed circles correspond to the system
with equal number of up- and down-spin electrons. The open
circles correspond to the spin-flip excitations.
}
\label{spectrum}
\end{center}
\end{figure}

From the exact diagonalization of the Hamiltonian we have 
obtained the energy spectra, $E_n$, and the corresponding wave
functions, $\Psi_n$, of the DNA system. The energy spectra for the
A-T and G-C base pairs are shown in Fig.~1. From this figure it
is clear that the excitation gap for the G-C base pairs,
$\Delta_{G-C} \approx 1.12$ eV, is larger than for the A-T
base pairs, $\Delta_{A-T} \approx 0.79$ eV. In both
cases the lowest excitation does not include electron spin-flip.
The spin-flip excitation (open circles) gap is equal to 
$\Delta_{G-C,spin}\approx 1.17$ eV for the G-C base pairs and  
$\Delta_{A-T,spin}\approx 0.94$ eV for the A-T pairs.
The main difference between the A-T and G-C pairs is that
for the G-C pairs the lowest spin-flip excitations are very close 
to the excitations without spin reversal. For example, for the G-C pairs 
the energy difference between these two types of
excitations is  $\Delta_{G-C,spin}-\Delta_{G-C}\approx 0.05$ eV,
which is smaller than corresponding value for the A-T pairs,  
$\Delta_{A-T,spin}-\Delta_{A-T}\approx 0.15$ eV.

It is interesting to compare the energy gaps of the interacting DNA
system to the corresponding gaps of the non-interacting system. It 
follows from Eq.~(1) that the energy gaps of the non-interacting 
system is equal to $\Delta_0 = \epsilon_l-\epsilon_h-2(t_l-t_h)$,
which gives $\Delta_{0,G-C} \approx 1.02$ eV for the G-C pairs and
$\Delta_{0,A-T} \approx 0.56$ eV for the A-T pairs. Comparing these 
values to the energy gaps of  corresponding interacting system we
conclude that both for G-C and for the A-T pairs we have interaction 
enhancement of the energy gaps by $0.1$ eV (for the G-C pairs) and by 
$0.23$ eV (for the A-T pairs). Again we see that the effect of 
interaction is more pronounced for the A-T pairs than for the 
G-C pairs. From all these results we conclude that the interaction 
has a weaker effect on the energy spectra for the G-C base pairs 
than for the A-T pairs.

In Ref.~\cite{yoo} the conductance of DNA molecules
was found to have activated dependence on temperature. 
Activation energies were extracted in Ref.~\cite{yoo}
to be 0.18 eV for A-T base pairs and 0.12 eV for G-C pairs at 
high temperatures. These activation energies are close to 
our interaction enhancement of the energy gaps as described
above. The activated nature of the DNA conductance means that 
electrons during their transport through a DNA molecule must 
overcome some potential barrier. Our results indicate that 
inter-electron interactions have considerable contribution 
to the activation energy of the electron transport. 

The numerically generated wave functions allow us to calculate the 
electron charge distribution along the DNA molecule from the equations

\begin{eqnarray}
\rho_{n,\uparrow}(k)&=&\sum_{i_1\ldots i_{N_{\uparrow}}}
         \sum_{j_1\ldots j_{N_{\downarrow}}}
   \left|\Psi_n \left(i_1,\ldots, i_{N_{\uparrow}}; j_1,\ldots,
j_{N_{\downarrow}}\right)\right|^2
\nonumber\\
&\times&\delta\left(k-i_1\right),
\nonumber \\
\rho_{n,\downarrow}(k)&=&\sum_{i_1\ldots
i_{N_{\uparrow}}}\sum_{j_1\ldots j_{N_{\downarrow}}}
   \left|\Psi_n\left(i_1,\ldots, i_{N_{\uparrow}}; j_1,\ldots,
j_{N_{\downarrow}}\right)\right|^2
\nonumber\\
&\times&\delta \left( k- j_1\right),
\end{eqnarray}                  
where $\rho_{n,\sigma}(k)$ is the density of the electrons with
spin $\sigma=\uparrow$ or $\downarrow$ in the state $n$ at the 
base pair $k$, and $\Psi_n\left(i_1,\ldots, i_{N_{\uparrow}}; 
j_1,\ldots, j_{N_{\downarrow}}\right)$ is the wave function of 
the state $n$ with $N_{\uparrow}$ electrons with spin
$\sigma =\uparrow$ and $N_{\downarrow}$ electrons with
spin $\sigma =\downarrow$. Here $i_1,\ldots, i_{N_{\uparrow}}$ and 
$j_1, \ldots, j_{N_{\downarrow}}$ are the coordinates of electrons in 
the base-pair representation. The results for the charge density
of the different base pairs are shown in Fig.~2 (for A-T) and in Fig.~3
(for G-C).

\begin{figure}
\begin{center}
\begin{picture}(120,130)
\put(0,0){\includegraphics{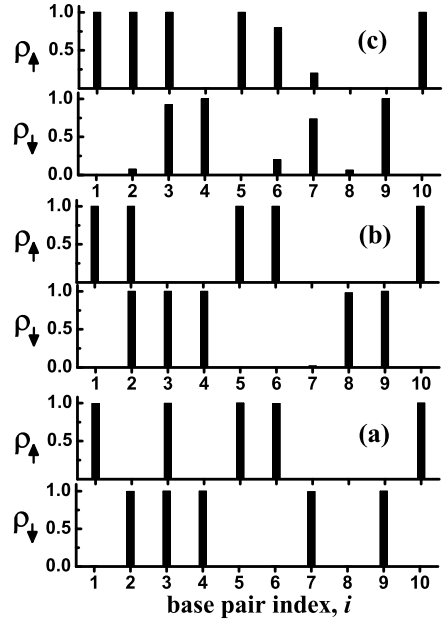}}
\end{picture}
\vspace*{3.0cm}
\caption{The density of electrons with up- and
down-spins  is shown for the A-T base pairs for (a) the
ground state; (b) the first excited state with equal
number of up- and down-spin electrons; (c) the first
spin-flip excited state.  
}
\label{densityAT}
\end{center}
\end{figure}

Clearly, for the A-T and the G-C pairs the charge distribution 
in the ground state is the same, which indicates that the electrons 
are strongly localized at the base pairs, i.e. at each base pair 
the electron density is equal to either 0 or 1 (the difference 
from 0 or 1 is less than 0.01). However, the charge distribution in 
the excited states of G-C and A-T systems show a different behavior 
(see Fig.~2b and Fig.~3b). For the G-C pairs, distribution of electrons 
with spin $\sigma=\downarrow $ (Fig.~3b) is the same as in the ground 
state (Fig.~3a). The excitation manifests itself only in the
redistribution of the electrons with spin  $\sigma=\uparrow$, making 
the single-electron states more delocalized. For the A-T pairs
the charge distribution in the excited state is different from the 
ground state for both spin directions (Fig.~2b). Contrary to the
case of the G-C pairs, here the single-electron states remain strongly 
localized even in the excited state.

\begin{figure}
\begin{center}
\begin{picture}(120,130)
\put(0,0){\includegraphics{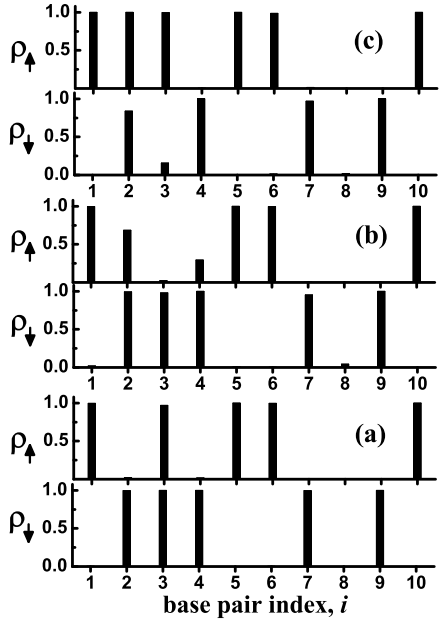}}
\end{picture}
\vspace*{3.0cm}
\caption{Same as in Fig.~2 but for the G-C base pairs.
}
\label{densityGC}
\end{center}
\end{figure}

The spin-flip excitations for the G-C and the A-T pairs also behave 
differently. For the A-T pairs the excitation is just spin-flip 
at one base (with base index $i=2$ in Fig.~2c) with a weak 
redistribution (delocalization) of other electrons over the base 
pairs. For the G-C pairs the spin-flip excitation corresponds to
spin-flip at one base pair with index $i=2$ (see Fig.~3c) with 
subsequent hopping of electron with spin $\sigma=\downarrow$ from 
base pair $i=3$ to the base pair $i=2$. Therefore, in the
case of the G-C pairs the many-particle spin-flip excitation is a
single-particle spin-flip hopping excitation. Since the electron-electron 
interactions tend to suppress the hopping processes, the difference 
in spin-flip excitations for the G-C and the A-T pairs also illustrate 
that the effect of electron interactions is less pronounced for
the G-C base pairs than for the A-T pairs.

In summary, we have performed theoretical calculations of the
electron energy spectrum, based on a two-leg charge ladder
model for the poly(dA)-poly(dT) DNA and poly(dG)-poly(dC) DNA
molecules. We take the electron-electron interactions and the 
electron spin degree of freedom fully into account in our model.
The energy spectra for the G-C and the A-T base pairs show a large gap
and the interaction was found to enhance the gap. The effect of
interaction is less pronounced for the G-C base pairs than that
of the A-T pairs. The spin-flip excitations are not the lowest
energy excitations. We also analyze the charge distribution for
the ground state as well as for the excitations. The present report
is the first step in our investigation of the electronic properties 
of the DNA. In our calculations of the energy spectra we have
not included the vibrational modes. These modes are very soft
in the DNA and can have strong effects on the excitation spectra, 
resulting in polaronic effects and strong renormalization of
electron-electron interactions. Such renormalization of 
inter-electron interactions should in turn have strong dependence 
on the temperature. This is because the temperature,
inducing the excitations of soft vibrational modes,
strongly affect the distances between the electrons
and as a result modify inter-electron interactions.
The effect of vibrational modes on the energy spectra and charge 
distribution of DNA molecules, as well as systems containing
more base pairs will be the subject of our future works.  

The work of one of us (T.C.) was supported by the Canada Research
Chair Program and the Canadian Foundation for Innovation Grant.


\begin{references}
\bibitem{note1}
No data is available for hopping integrals for interchain 
transport, but estimated to be $t_{hl}\approx 0.01$ for GC and AT.
\bibitem{note2}
Only Hertree terms are available \protect\cite{brunaud}.
\end{references}
\end{document}